\documentstyle[preprint,eqsecnum,aps]{revtex}
\begin{document}
\draft
\preprint{KYUSHU-HET-22}
\title{Six-body Light-Front Tamm-Dancoff approximation\\
 and wave functions
for the massive Schwinger model}
\author{
Koji Harada\footnote{e-mail address:
koji1scp@mbox.nc.kyushu-u.ac.jp},
Atsushi Okazaki\footnote{e-mail address:
zaki1scp@mbox.nc.kyushu-u.ac.jp},
and
Masa-aki Taniguchi\footnote{e-mail address:
mass1scp@mbox.nc.kyushu-u.ac.jp}
}
\address{
Department of Physics, Kyushu University\\
Fukuoka 812, JAPAN
}
\date{February 1995}
\maketitle
\begin{abstract}
The spectrum of the massive Schwinger model in the strong coupling
region is obtained by using the light-front Tamm-Dancoff (LFTD)
approximation up to including six-body states.
We numerically confirm that the two-meson bound state has a
negligibly small six-body component. Emphasis is on the usefulness
of the information about states (wave functions). It is used for
identifying the three-meson bound state among
the states below the three-meson threshold.
We also show that the two-meson bound state
is well described by the wave function of the relative motion.
\end{abstract}

\pacs{11.10.Kk,11.10.St,11.15.Tk}

\narrowtext

%%%%%%%%%%%%%%%%%%%%%%%%%%%%%%%%%%%%%%%%%%%%%%%%%%%%%%%%%%%%%%%%%
% section Introduction
%%%%%%%%%%%%%%%%%%%%%%%%%%%%%%%%%%%%%%%%%%%%%%%%%%%%%%%%%%%%%%%%%
\section{Introduction}\label{intro}

In a previous paper\cite{hsty}, we investigated the massive
Schwinger model\cite{cjs,coleman} with
$SU(2)_f$ in the light-front Tamm-Dancoff (LFTD)
approximation\cite{td,lftd} up to
including four-body states. We showed, by examining the wave
functions, that the lightest isosinglet state can be regarded as
a bound state of two ``pions.'' This observation naturally led
us to the answer to the question raised by Coleman\cite{coleman}
why it is so light. The LFTD approximation has
been proved to be one of the most powerful non-perturbative methods
to date in the investigation of relativistic bound states, at least
in two dimensions, although we have to face the difficult
renormalization problem in higher dimensions.

The validity of the LFTD approximation is based on the plausible
hope that the sea quark/gluon contributions are small in the
light-cone quantization because pair creations/annihilations are
suppressed\cite{bmpp}. Typically, the lightest particles are
expected to be in the valence states.
It is generally true in the models so far investigated. The above
mentioned state (the bound state of two ``pions'') is an important
exception. With this exception, one might think that such a state
would have non-negligible many-body components too.
It is one of our purposes of this paper to show numerically that
it is unlikely by examining the single-flavor model.

We also investigate the three-meson bound state
of the single-flavor model. Its existence has been discussed by
Coleman\cite{coleman} by using the bosonization
technique. He showed that, in the strong coupling limit with the
zero vacuum angle, there exists a stable three-meson bound state
and it is unstable when the vacuum angle is non-zero. We look for
a candidate which can be interpreted as a bound state of three
mesons in our numerical results.

In order to investigate these problems, it is necessary to do LFTD
calculations up to including six-body states.
Such calculations are very hard without any technical refinement.
To make these calculations feasible we have made two points:
(1) We take a simple set of basis
functions in order to reduce CPU time. A clever choice of basis
functions is essential as in quantum chemistry calculations.
Note that our choice of basis functions in a finite domain will
be also useful for higher dimensions. Even in higher dimensions
the longitudinal momenta $p^+_i$ of constituents are restricted
to a finite domain $0\le p^+_i\le P^+$ with $\sum_i p^+_i=P^+$,
where $P^+$ is the total momentum. (2) The
three-meson  bound state, if it exists, must be in the continuum
unless it is lighter than two mesons, and is therefore apparently
difficult to find.  We can however find a candidate among several
states by looking at the wave functions. The points are
that a three-meson state must be charge conjugation odd
and that below
the three-meson threshold, six-body components should be very
small except for three-meson bound states. A more detailed
discussion is given in Sec.\ \ref{results}.

We emphasize that the information about states (wave functions) is
very useful. It is used for identifying the three-meson
bound state, as is said above. As another example, we introduce
the wave function of the
relative motion of the two-meson bound state and
try to describe the
bound state in terms of the wave function. Although the concept of
``relative motion'' of a relativistic bound state is somewhat
awkward, we however find that the two-meson bound state
is well described
in terms of the wave function of the relative motion,
in the sense that a
smaller set of basis functions motivated by the
concept of the relative
motion gives a good approximation. It gives us
a qualitative picture of
the bound state.

 We summarize the results: (1) The masses of
the lowest states do not change even if we include
six-body states. (2)
In particular, the state which can be regarded as
a bound state of two
mesons has a negligible six-body component.
(3) We find a candidate for
the bound state of three mesons. (4) The wave
function of the relative motion of the two-meson
bound state describes the bound state well. We can have a
picture that in
the strong coupling region it is loosely bound,
while in the weak
coupling region it is tightly bound, compare
to the size of the meson.

The massive Schwinger model\cite{schwinger,ls}
has been discussed by many
authors in the light-cone quantization.
Bergknoff\cite{bergknoff} did the
first LFTD calculations. Mo and Perry\cite{moperry}
refined his
calculations by the use of basis functions.
Their calculations include
only up to four-body states. Eller, Pauli,
and Brodsky\cite{epb,yh}
discussed the massless and the massive Schwinger models in the
discretized light cone quantization (DLCQ).
Our work is based especially
on the papers by Bergknoff, and Mo and Perry.
We try to keep our notation
as close as possible to that of our previous paper\cite{hsty}.

In Sec. \ref{theory}, we present basic facts
and formulas on the massive
Schwinger model to make this paper self-contained.
The model is quantized
on the light cone and the Tamm-Dancoff truncation is
made up to including
six-body states. The wave functions are expanded
in terms of a new set of
basis functions. The numerical results are shown in
Sec. \ref{results}. We identify two-meson and
three-meson bound states.
The two-meson bound state is shown to have a
negligibly small six-body
component. The three-meson bound state is charge conjugation
odd and has a
large six-body component
compare to those of other states below the
three-meson threshold.
In Sec. \ref{meson}, we introduce a meson
operator which (approximately)
creates a meson from the vacuum. By using
the meson operator, we also
introduce the wave function of the relative motion.
Sec. \ref{summary} is devoted to discussions.

%%%%%%%%%%%%%%%%%%%%%%%%%%%%%%%%%%%%%%%%%%%%%%%%%%%%%%%%%%%%%%%%
% section formulation
%%%%%%%%%%%%%%%%%%%%%%%%%%%%%%%%%%%%%%%%%%%%%%%%%%%%%%%%%%%%%%%%
\section{Formulation}\label{theory}

\subsection{Definition of the model}\label{define}

The massive Schwinger model\cite{cjs,coleman} is
two-dimensional QED with
a massive fermion. It is not exactly solvable in
contrast to the massless
one\cite{schwinger,ls}. The Lagrangian is given by
\begin{equation}
	{\cal L}=-{1\over4}F_{\mu\nu}F^{\mu\nu}
	+\bar\psi\left[\gamma^{\mu}
	\left(i\partial_{\mu}-eA_{\mu}\right)
	-m\right]\psi\label{lagrangian}\ ,
\end{equation}
where $F_{\mu\nu}=\partial_{\mu}A_{\nu}-
\partial_{\nu}A_{\mu}$. In
two dimensions, the coupling constant $e$
has mass dimension. It is
therefore useful to measure all dimensionful
quantities in units of
$e/\sqrt{\pi}$. We hereafter set $e/\sqrt{\pi}=1$.
Strong couplings
correspond to small fermion masses.

In the light-cone gauge ($A^+=0$),
the  only
independent variable is
$\psi_R$ in the light-cone quantization. $A^-$
and $\psi_L$ are
expressed in terms of $\psi_R$ as follows:
\begin{eqnarray}
	A^- &=& \sqrt{\pi}{1\over (i\partial_{-})^2}j^+,
	\label{A^-} \\
    \psi_L &=& {m\over\sqrt{2}}
    {1\over i\partial_{-}}\psi_R \label{psi_L},
\end{eqnarray}
with $j^{+}(x^{-})=\sqrt{2}:\psi_{R}^{\dag}
(x^{-})\psi_{R}(x^{-}):$. We use the principal value
prescription for $(i\partial_-)^{-1}$ and
$(i\partial_-)^{-2}$ as in Refs.\ \cite{hsty,moperry}.

Eliminating $A^-$ and $\psi_L$ by using (\ref{A^-}) and
(\ref{psi_L}), one obtains the light-cone Hamiltonian $P^-$,
\begin{eqnarray}
	& &P^{-}=P_{free}^{-}+P_{int}^{-}\label{hamiltonianpsi}\ ,
	\nonumber\\
	& &P_{free}^{-}={m^2\over2\sqrt{2}}
	\int_{-\infty}^{\infty}dx^{-}\psi_{R}^{\dag}(x^{-})
	{1\over i\partial_{-}}\psi_{R}(x^{-})\ ,\\
	& &P_{int}^{-}={\pi\over 2}\int_{-\infty}^{\infty}dx^{-}
	j^{+}(x^{-}){1\over (i\partial_{-})^2}
	j^{+}(x^{-})\ .\nonumber
\end{eqnarray}

We expand $\psi_{R}$ in terms of the creation and
annihilation operators,
\begin{equation}
       \psi_{R}(x^{-})={1\over 2^{1/4}}\int_{0}^{\infty}
       {dk^{+}\over (2\pi)\sqrt{k^{+}}}\left[
       b(k^{+})e^{-ik^{+}x^{-}}+d^{\dag}(k^{+})
       e^{ik^{+}x^{-}}\right]\ ,
\end{equation}
where $b(k^+)$ and $d(k^+)$ satisfy the following
anti-commutation
relations,
\begin{equation}
	\left\{ b(k^+),b^{\dag}(l^+)\right\}=
	\left\{ d(k^+),d^{\dag}(l^+)\right\}=
	(2\pi)k^{+}\delta(k^+-l^+)\ ,
\end{equation}
derived from
$\left\{\psi_{R}(x^-),\psi_R^{\dag}(y^-)\right\}
=(1/\sqrt{2})\delta(x^--y^-)$.
One may express $P^-$ entirely in terms of $b(k^+)$
and $d(k^+)$ (and
their Hermitian conjugates). We refer the reader to
Ref.\ \cite{moperry}
for the explicit form.

We work in a truncated Fock space in which a state with
total light-cone
momentum $P^+={\cal P}$ is expressed as
\begin{eqnarray}
|\psi\rangle_{\cal P}&=&
        |2\rangle_{\cal P}+|4\rangle_{\cal P}+
        |6\rangle_{\cal P}\ ,\label{states}\\
|2\rangle_{\cal P}&=&\int_{0}^{{\cal P}}{dk_1dk_2\over
	\sqrt{(2\pi)^2k_1k_2}}\delta(k_1+k_2-{\cal P})
	\psi_{2}(k_1,k_2)
	b^{\dag}_1d^{\dag}_2|0\rangle\ ,\nonumber\\
|4\rangle_{\cal P}&=&{1\over 2}\int_{0}^{{\cal P}}
        \prod_{i=1}^{4}{dk_{i}\over \sqrt{(2\pi)k_{i}}}
	\delta(\sum_{i=1}^{4}k_{i}-{\cal P})
	\psi_{4}(k_{1},k_{2};k_{3},k_{4})
	b^{\dag}_{1}b^{\dag}_{2}d^{\dag}_{3}d^{\dag}_{4}|0\rangle\ ,
	\nonumber\\
|6\rangle_{\cal P}&=&{1\over 3!}\int_{0}^{{\cal P}}
        \prod_{i=1}^{6}{dk_{i}\over \sqrt{(2\pi)k_{i}}}
	\delta(\sum_{i=1}^{6}k_{i}-{\cal P})
	\psi_{6}(k_{1},k_{2},k_{3};k_{4},k_{5},k_{6})
	b^{\dag}_{1}b^{\dag}_{2}b^{\dag}_{3}
	d^{\dag}_{4}d^{\dag}_{5}d^{\dag}_{6}|0\rangle\
	,\nonumber
\end{eqnarray}
where we use the abbreviated notations,
$b^{\dag}_{i}=b^{\dag}(k_{i})$,
$d^{\dag}_{i}=d^{\dag}(k_{i})$. We rescale momenta,
$k_i\rightarrow x_i=k_i/{\cal P}$ and the
wave functions, $\psi_{2}(k_{1},k_{2})$,
$\psi_{4}(k_{1},k_{2};k_{3},k_{4})$,
and $\psi_{6}(k_{1},k_{2},k_{3};k_{4},k_{5},k_{6})$ are
replaced by $\psi_{2}(x_{1},x_{2})$,
${\cal P}^{-1}\psi_{4}(x_{1},x_{2};x_{3},x_{4})$,
and
${\cal P}^{-2}\psi_{6}(x_{1},x_{2},x_{3};x_{4},x_{5},x_{6})$,
respectively.

The wave functions $\psi_{2}$, $\psi_{4}$
and
$\psi_{6}$ must satisfy the following symmetry properties
due to Fermi
statistics,
\begin{eqnarray}
	\psi_{4}(x_{1},x_{2};x_{3},x_{4})&=&
	-\psi_{4}(x_{2},x_{1};x_{3},x_{4})
	=-\psi_{4}(x_{1},x_{2};x_{4},x_{3})=
	\psi_{4}(x_{2},x_{1};x_{4},x_{3})\ ,
	\nonumber\\
	\psi_{6}(x_{1},x_{2},x_{3};x_{4},x_{5},x_{6})&=&
	-\psi_{6}(x_{2},x_{1},x_{3};x_{4},x_{5},x_{6})
	=-\psi_{6}(x_{1},x_{3},x_{2};x_{4},x_{5},x_{6})\\
	&=&-\psi_{6}(x_{1},x_{2},x_{3};x_{5},x_{4},x_{6})=
	-\psi_{6}(x_{1},x_{2},x_{3};x_{4},x_{6},x_{5})
	\ etc.\ .\nonumber
\end{eqnarray}
If we require that this state has a definite property
under charge
conjugation transformation, we have further
conditions on these wave
functions,
\begin{eqnarray}
	& &\psi_{2}(x_{1},x_{2})=\mp\psi_{2}(x_{2},x_{1})\ ,
	\nonumber\\
	& &\psi_{4}(x_{1},x_{2};x_{3},x_{4})=
	        \pm\psi_{4}(x_{3},x_{4};x_{1},x_{2})\ ,
	        \label{ccs}\\
	& &\psi_{6}(x_{1},x_{2},x_{3};x_{4},x_{5},x_{6})=
	        \mp\psi_{6}(x_{4},x_{5},x_{6};x_{1},x_{2},x_{3})\ .
\nonumber
\end{eqnarray}
The upper/lower sign in (\ref{ccs}) corresponds to
charge conjugation
even/odd.

The Einstein-Schr\"odinger equation $M^2|\psi\rangle_{\cal
P}=2P^{-}P^{+}|\psi\rangle_{\cal P}$ leads to a set of complicated
eigenvalue equations for the wave functions. It can be converted to a
single matrix eigenvalue problem by expanding the wave functions in terms
of basis functions, which we discuss in the next subsection.

\subsection{Basis functions}\label{basis}
It has been known that the wave function
$\psi_2(x,1-x)$ behaves as
$x^\beta$ in the vicinity of $x=0$\cite{bergknoff},
with $\beta$ being the
solution of the equation
$m^2-1+\pi\beta \cot(\pi\beta)=0$. By taking it
into account, Mo and Perry concluded that a
useful choice of the basis
functions for the wave functions is given in terms of
Jacobi polynomials,
$P^{(\beta,\beta)}_n$. In a previous paper\cite{hsty},
we propose a simpler
set of basis functions, essentially equivalent
to that of Mo and Perry.
We now propose another set of basis functions
which leads to a
drastic reduction of CPU time. We expand
the wave functions as follows.
\begin{eqnarray}
	& &\psi_{2}(x,1-x)=\sum_{k=0}^{N_{2}}a_{k}F_{k}(x,1-x)\ ,
	\nonumber\\
	& &\psi_{4}(x_{1},x_{2};x_{3},x_{4})=
	\sum_{{\bf k}}^{N_{4}}b_{{\bf k}}
	G_{{\bf k}}(x_{1},x_{2};x_{3},x_{4})\ ,\
	\sum_{i=1}^4x_i=1\ ,\\
	& &\psi_{6}(x_{1},x_{2},x_{3};x_{4},x_{5},x_{6})=
	        \sum_{{\bf K}}^{N_{6}}c_{{\bf K}}
	H_{{\bf K}}(x_{1},x_{2},x_{3};x_{4},x_{5},x_{6})\ ,\
	        \sum_{i=1}^6x_i=1\ ,\nonumber
\end{eqnarray}
where we use the following basis functions:
\begin{equation}
	F_{k}(x,1-x)=
	\left\{\begin{array}{l}\left[x(1-x)\right]^{\beta+k}\\
	\left[x(1-x)\right]^{\beta+k}(2x-1)\ ,\end{array}\right.
	\label{two}
\end{equation}
\begin{equation}
	G_{{\bf k}}(x_{1},x_{2};x_{3},x_{4})=
	(x_{1}x_{2}x_{3}x_{4})^{\beta}(x_{1}x_{2})^{k_{1}}
	(x_{3}x_{4})^{k_{2}}(x_{1}-x_{2})(x_{3}-x_{4})
	(x_{1}+x_{2})^{k_{3}}\
	,\label{four}
\end{equation}
\begin{eqnarray}
	&&H_{{\bf K}}(x_{1},x_{2},x_{3};x_{4},x_{5},x_{6})
	=(x_{1}x_{2}x_{3}x_{4}x_{5}x_{6})^{\beta}\nonumber\\
	&&\qquad\qquad
	\times\left[(x_{1}x_{2})^{k_{1}}(x_{1}+x_{2})^{k_{2}}(x_{1}-x_{2})
	+(1,2,3\;\;cyclic)\right]\nonumber\\
	&&\qquad\qquad
	\times\left[(x_{4}x_{5})^{k_{3}}(x_{4}+x_{5})^{k_{4}}
	(x_{4}-x_{5})
	+(4,5,6\;\;cyclic)\right]
	(x_{1}+x_{2}+x_{3})^{k_{5}}\ .\label{six}
\end{eqnarray}
We abbreviate the upper limits of the sums. In reality,
$N_2=2(M_1+1)$,
and it means that $k$ in (\ref{two}) runs from $0$ to
$M_1$. Similarly,
$N_4=(M_2+1)^2(M_3+1)$, i.e., $k_1,k_2=0,\cdots,M_2$,
$k_3=0,\cdots,M_3$ in (\ref{four})
and $N_6=(M_4+1)^2(M_5+1)^2(M_6+1)$, i.e.,
$k_1,k_3=1,\cdots,M_4+1$, $k_2,k_4=0,\cdots,M_5$, and
$k_5=0,\cdots,M_6$ in
(\ref{six}).
The important point in choosing this set of basis functions
is to reduce the number of the factors of the type
$(x_1+x_2)^k$. This reduction allows us to express
the basis functions in
a simple way in the source code. For example, $G_{\bf
k}(x_1,x_2;x_3,x_4)$ may be written as
\begin{equation}
	G_{\bf k}(x_1,x_2;x_3,x_4)=
	\sum^{k_3}_{j=0}\sum^3_{i=0}(-1)^{2-i_1-i_2}
	\pmatrix{k_3\cr j}(x_1x_2x_3x_4)^\beta
	x_1^{N_1}x_2^{N_2}x_3^{N_3}x_4^{N_4},
\end{equation}
with $N_1=k_1+i_1+j$, $N_2=k_1+1-i_1+k_3-j$, $N_3=k_2+i_2$,
$N_4=k_2+1-i_2$, and $i=2i_1+i_2$. (We use a
binary number for $i$.)
Without using this new set of basis functions, six-body
LFTD calculations would be much more heavy.

We have explicitly separated the two-body basis
functions into charge
conjugation eigenfunctions. But we have not done
that for four-body and
six-body basis functions because it makes the
expressions so complicated
that the drastic reduction of CPU time cannot
be expected. We determine
the charge conjugation property of an eigenstate
by looking at the
two-body state. From our experience, we know
that it is a reliable way.

With these expansions, the Einstein-Schr\"odinger
equation becomes a
(generalized) matrix eigenvalue problem, which can
be solved numerically. Calculations of the matrix
elements can be carried out analytically
by using the formulas
(and their generalizations) collected in an appendix
of Ref.\ \cite{hsty}.

%%%%%%%%%%%%%%%%%%%%%%%%%%%%%%%%%%%%%%%%%%%%%%%%%%%%%%%%%%%%%%%%
% section numerical results
%%%%%%%%%%%%%%%%%%%%%%%%%%%%%%%%%%%%%%%%%%%%%%%%%%%%%%%%%%%%%%%%
\section{Numerical results}\label{results}

\subsection{Convergence}\label{conv}

First of all we have to see how many basis functions
are enough to produce
reliable results. We set $m=0.1$ and gradually
increase the number of
basis functions. Fig.\ \ref{twoTD} shows the lowest
mass states in the  calculation
including only two-body basis functions. The lowest
state is the meson
state. It is charge conjugation odd. Its mass is
$1.18160$ at $N_2=10\
(M_1=4)$. The dashed line indicates the two-meson
threshold. Note that
the convergence is good enough for $N_2=10$.

As we increase the number of four-body basis functions,
(keeping $N_2=10$)
a state goes down
below the two-meson threshold as shown in
Fig.\ \ref{fourTD}. We regard this state as
the two-meson bound state. On the other hand,
the lowest state, the meson,
is little
affected by the inclusion of the four-body states.
Its mass is $1.18103$
at $N_4=80\ (M_2=3, M_3=4)$,
decreased only $0.05\%$. It is due to a negligibly
small four-body
component, $0.002\%$. Note that all the state above
the two-meson threshold
go down as $N_4$ increases. We find that $N_4=80$ is
enough for the
convergence for the lightest two states.

We proceed to the six-body calculations, keeping
$N_2=10$ and $N_4=80$ $(N_2+N_4=90$).
As seen in Fig.\ \ref{sixTD}, the convergence is
quite good for the states below the
three-meson threshold indicated by the dotted line.
The three-meson threshold is given by the sum of the
meson mass and the mass of
the two-meson bound state. Again, the meson is not
affected by the inclusion of
six-body states. The two-meson bound state does not
change either. The
mass is $2.30980$ at $N_6=36\ (M_4=2, M_5=1, M_6=0)$
which should be
compared to $2.31004$ in the four-body calculations.
The state just below
the three-meson threshold can be regarded as the
three-meson bound state,
as we discuss shortly. It seems that
$N_6=36$ $(N_2+N_4+N_6=126)$ is enough for the convergence
for the lightest states. In the
following, we restrict ourselves to this case.

\subsection{Two-meson bound state}\label{2meson}

The second lightest state can be regarded as a bound
state of two mesons.
Its mass is $2.30980$ at $m=0.1$. It has a $72.825\%$
two-body component,
$27.174\%$ four-body component and $0.001\%$ six-body
component. The
ratios change as the fermion mass changes. The smaller
gets the fermion
mass, the larger four-body component it has. For
example, at $m=0.01$, it
has $54.408\%$ two-body component, $45.592\%$ four-body
component and $0.0004\%$
six-body component.
It is important to notice that it has a negligibly
small six-body component. From this result we presume that it will not have
any many-body components
even if we could include higher Fock states. This
state is charge
conjugation even.

\subsection{Three-meson bound state}\label{3meson}

We identify the three-meson bound state by the
following criteria; (1)
Its mass must be below the three-meson threshold.
(2) It must have a large six-body component relative
to the other states below
the three-meson threshold, at least in the strong
coupling region. (3) It must be charge conjugation odd.

The first criterion is a trivial one, and is necessary for
distinguishing it from three-meson scattering states.

The second criterion is based on the observation that the meson is
almost completely in the valence state in the strong coupling region. As
we discuss in detail in the next section, one may consider an
(approximate) meson creation operator $A^\dagger$. Thus a
three-meson
state may be represented as
$\sim (A^\dagger)^3|0\rangle$, which implies that the
three-meson bound
state has a large six-body component and negligibly small
many-body components.
Similarly a two-meson state has a negligibly small six-body
component.

The third criterion comes from the fact that a meson state
is charge conjugation odd.

We find such a state that satisfies all of these criteria.
Its mass is $3.39181$ at $m=0.1$, well below the threshold,
$3.49083$.
It has a $65.170\%$ two-body component,
$34.292\%$ four-body components and $0.538\%$
six-body component. All
states near this have smaller six-body components,
typically a few hundredths percent or less.
For smaller fermion masses, the six-body component
of the three-meson bound state become larger. For
example, at $m=0.001$, the two-, four-, and six-body
components are $42.013\%$, $55.013\%$, and $2.974\%$
respectively. It is charge conjugation odd.

One might be surprised that it has a small
six-body component, and might suspect that it is a two-meson
state. In Sec. \ref{summary}, we will argue that it cannot be
regarded as a two-meson state.

%%%%%%%%%%%%%%%%%%%%%%%%%%%%%%%%%%%%%%%%%%%%%%%%%%%%%%%%%%%%%%%%
% section wave function
%%%%%%%%%%%%%%%%%%%%%%%%%%%%%%%%%%%%%%%%%%%%%%%%%%%%%%%%%%%%%%%%
\section{Wave function for the relative motion of the
two-meson bound state}\label{meson}

In the previous section, we utilize the information
of the wave function
to identify the three-meson bound state. It is an
outstanding feature
that we can get such information. In this section,
we consider another
example in which the information of the wave
function is crucial. We
introduce a meson creation operator to have a
qualitative picture of the two-meson bound
state by considering the wave function of the relative motion.

\subsection{Meson operator}\label{operator}
Let us introduce an operator $a^\dagger(p)$,
\begin{eqnarray}
	a^{\dag}(p)
	%&=&\int_{-\infty}^{\infty}dxj^+(x)e^{-ipx}\ (p>0)\\
	&=&\int_{0}^{p}{dk\over (2\pi)\sqrt{k(p-k)}}
	    b^{\dag}(k)d^{\dag}(p-k)\nonumber\\
	&+&\int_{0}^{\infty}{dk\over (2\pi)\sqrt{k(p+k)}}
	\left[b^{\dag}(p+k)b(k)-d^{\dag}(p+k)d(k)\right]\ .
	\label{masslessmeson}
\end{eqnarray}
It is easy to show that it satisfies the following commutation
relations,
\begin{eqnarray}
	&&\left[a(p),a^{\dag}(q)\right]=p\delta(p-q)\ ,
	                \label{mesoncommutator}\\
	&&\left[a(p),a(q)\right]=
	\left[a^{\dag}(p),a^{\dag}(q)\right]=0\ ,
	                \nonumber
\end{eqnarray}
where $a(p)$ is the Hermitian conjugate to  $a^\dagger(p)$ and
annihilates the vacuum,
\begin{equation}
	a(p)|0\rangle=0.
\end{equation}
By using these operators,
the Hamiltonian can be written in the following form,
\begin{equation}
	P^-={m^2\over 4\pi}\int_{0}^{\infty}{dk\over k^2}
	     \left[b^{\dag}(k)b(k)+d^{\dag}(k)d(k)\right]
	     +{1\over 2}\int_{0}^{\infty}{dp\over p^2}
	     a^{\dag}(p)a(p)\ .
\end{equation}
Note that in the massless case $m=0$, the
Hamiltonian is diagonal and the
eigenstates are the Fock states of $a^\dagger$. The operator
$a^\dagger$ is the creation operator of the meson
in the (massless)
Schwinger model, which is equivalent to a
free massive scalar theory,
\begin{equation}
	|\mbox{meson}(m=0)\rangle_{{\cal P}}=
	a^{\dag}({\cal P})|{0}\rangle
	=\int_{0}^{\cal P}{dk\over (2\pi)\sqrt{k({\cal P}-k)}}
	b^{\dag}(k)d^{\dag}({\cal P}-k)|{0}\rangle.
\end{equation}
The meson is structureless in the sense that the wave
function has no
momentum dependence, $\psi(k,{\cal P}-k)=1$ for any $k$.
Compare with
(\ref{mesonoperator}) below.

Once the fermion mass is introduced, the Fock
states of $a^\dagger$ are
no longer the eigenstates of the Hamiltonian.
The results in the previous
section suggest, however, that one may introduce
an approximate meson
creation
operator whose Fock states are approximate
eigenstates of the Hamiltonian.
We have seen that the wave function of the
meson behaves as
$\psi\sim[x(1-x)]^{\beta}$ \cite{bergknoff} and the higher
Fock components are negligible
in the strong coupling region. Taking into
account these things, we
propose the following approximate meson creation operator,
\begin{eqnarray}
	A^{\dag}(p)&=&\int_{0}^{p}{dk\over (2\pi)\sqrt{k(p-k)}}
	        \psi(k,p-k)b^{\dag}(k)d^{\dag}(p-k)
	                                  \label{mesonoperator}\\
	&+&\int_{0}^{\infty}{dk\over (2\pi)\sqrt{k(p+k)}}
	\varphi(p+k,k)
	\left[b^{\dag}(p+k)b(k)-d^{\dag}(p+k)d(k)\right]\ ,
	\nonumber
\end{eqnarray}
where $\psi$ should be equivalent to $\psi_2$ for the
meson in the previous section and
therefore had been known numerically,
\begin{equation}
|\mbox{meson}(m\ne0)\rangle_{\cal P}\approx
A^\dagger({\cal P})|0\rangle
=\int^{{\cal P}}_0{dk\over (2\pi)\sqrt{k({\cal P}-k)}}
\psi(k,{\cal P}-k)b^\dagger(k)d^\dagger({\cal P}-k)|0\rangle.
\label{approxmeson}
\end{equation}
Compare with (\ref{states}).
The shape of
$|\psi|^2$ is shown in
Fig.\ \ref{mesonWF}. On the other hand, $\varphi$ cannot
be determined by looking at the meson state. But
it affects two-meson
states and can be determined by examining the
two-meson bound state, at
least in principle.

It is hard to estimate the errors of the approximate
{\it operator} $A^\dagger(p)$, though the {\it state}
(\ref{approxmeson})
has been shown to be a fairly good approximation.
For small fermion masses (small $\beta$), we expect that
$\psi=
1+{\cal O}(\beta)$ and $\varphi=1+{\cal O}(\beta)$,
and the errors are expected to be ${\cal O}(\beta)$ because
$A^\dagger(p)$ reduces to $a^\dagger(p)$ for $\psi=\varphi
\equiv1$.

\subsection{Wave function of the relative motion}
\label{WFRM}
Let us attempt to describe the two-meson bound
state by introducing the
wave function of the relative motion.
Such a description is based on the
{\it assumption\/} that it is a {\it two-meson\/} state,
i.e., that the two
mesons in the bound state would not be
distorted too much. The assumption
is justified {\it a posteriori\/} in the
strong coupling region.

Under this assumption, the two-meson
bound state may be written as
\begin{eqnarray}
	|{\mbox{two-meson}}\rangle_{\cal P}&\equiv&
	{1\over\sqrt{2}}
	\int_{0}^{\cal P}{dp_{1}dp_{2}\over\sqrt{p_1p_2}}
	\delta(p_{1}+p_{2}-{\cal P})
	\Phi(p_{1},p_{2})A^{\dag}(p_{1})A^{\dag}(p_{2})|{0}\rangle
	                                     \label{twomeson}\\
	&=&\int_{0}^{\cal P}{dk\over (2\pi)\sqrt{k({\cal P}-k)}}
	      \Psi_{2}(k,{\cal P}-k)b^{\dag}(k)
	d^{\dag}({\cal P}-k)|{0}\rangle\nonumber\\
	&+&{1\over 2}\int_{0}^{\infty}
	\prod_{i=1}^{4}{dk_{i}\over \sqrt{(2\pi)k_{i}}}\delta(
	\sum_{i=1}^{4}k_{i}-{\cal P})
	\Psi_{4}(k_{1},k_{2};k_{3},k_{4})b^{\dag}_1b^{\dag}_2
	d^{\dag}_3d^{\dag}_4|{0}\rangle\ ,\nonumber
\end{eqnarray}
where we have substituted (\ref{mesonoperator}).
The wave functions
$\Psi_2$ and $\Psi_4$ are expressed in terms of
$\psi$ and $\varphi$ in
the following way,
\begin{eqnarray}
	\Psi_{2}(k_{1},k_{2})&=&
	{1\over\sqrt{2}}\int_{0}^{k_{1}}
	{dq\over\sqrt{q({\cal P}-q)}}
	\Phi(q,{\cal P}-q)
	\varphi(k_{1},k_{1}-q)\psi(k_{2},k_{1}-q)
	\label{Psi2}\\
	&-&{1\over\sqrt{2}}\int_{0}^{k_{2}}
	{dq\over\sqrt{q({\cal P}-q)}}
	\Phi(q,{\cal P}-q)
	\varphi(k_{2},k_{2}-q)\psi(k_{1},k_{2}-q)\ ,
	\nonumber\\
	\Psi_{4}(k_{1},k_{2};k_{3},k_{4})&=&
	-{\Phi(k_{1}+k_{3},k_{2}+k_{4})\over
	\sqrt{2(k_1+k_3)(k_2+k_4)}}
	       \psi(k_{1},k_{3})\psi(k_{2},k_{4})
	\label{Psi4}\\
	&+&{\Phi(k_{1}+k_{4},k_{2}+k_{3})\over
	\sqrt{2(k_1+k_4)(k_2+k_3)}}
	       \psi(k_{1},k_{4})\psi(k_{2},k_{3})\ .
	\nonumber
\end{eqnarray}
The wave function $\Phi$ is that
of the relative motion of
the two mesons in the two-meson bound state.

The meson operator (\ref{mesonoperator}) does
not exactly satisfy the same commutation relations as
(\ref{mesoncommutator}), but only approximately.
Thus $\Phi(p_1,p_2)$ does
not need to be symmetric under the exchange of $p_1$ and $p_2$.
Nevertheless we regard it as being symmetric throughout
this paper. Note
that the ansatz (\ref{twomeson}) is consistent with the
charge conjugation
symmetry, that is, $\Psi_2(k_1,k_2)$ is antisymmetric
in $k_1$ and
$k_2$, $\Psi_4(k_{1},k_{2};k_{3},k_{4})$ is symmetric
under the exchange
of $(k_1,k_2)$ and $(k_3,k_4)$.

It is interesting to note that this ansatz drastically
reduces the
degrees of freedom in the functional space.
Due to the assumption that the meson wave function
would not
be distorted too much in the bound state, only one
degree of freedom,
i.e., the relative motion of the mesons,
comes in. It seems natural to expand $\Phi(x_1,x_2)$
(symmetric in $x_1$
and $x_2$) as
\begin{equation}
	\Phi(x_1,x_2)=
	\sum_{l=0}^{N}B_l[x_1x_2]^{l+{1\over2}}\ ,
	x_1+x_2=1\ ,
	\label{relwf}
\end{equation}
where $B_l$ is a coefficient to be determined numerically.
Taking into account the fact that $\psi(x_1,x_2)$ is
well approximated by
$a_0(x_1x_2)^\beta$, with $a_0$ being the
normalization constant,
$a_0=[B(2\beta+1,2\beta+1)]^{-1/2}$ ($B$ is a Beta function),
one may consider the following basis function expansions,
\begin{eqnarray}
	\Psi_2(x,1-x)&=&
	\sum_{l=0}^{N}A_l[x(1-x)]^{\beta+l}(2x-1)\ ,
	\label{reltwo}\\
	\Psi_4(x_1,x_2,x_3,x_4)&=&
	\sum_{l=1}^{N}{B_la_0^2\over\sqrt{2}}
	(x_1x_2x_3x_4)^{\beta}\label{relfour}\\
	\nonumber
	&&\times\left\{-[(x_1+x_3)(x_2+x_4)]^l
	+[(x_1+x_4)(x_2+x_3)]^l\right\}\ ,
\end{eqnarray}
where $A_l$ is another coefficient to be determined numerically.
Note that this set of basis functions is much
simpler than the original
one (\ref{four}).

We calculate the mass and the wave functions (i.e., the
coefficients $A_l$ and $B_l$)
of the two-meson bound state by using this set of basis functions
with $N=7$. The mass is calculated as
$2.04180$ at $m=0.01$. This result is surprisingly
good for this small set of basis functions. It is even better
than the result of our full-set calculations,
$m=2.05612$. This is because of the factors like
$[(x_1+x_3)(x_2+x_4)]^l$ in (\ref{relfour}),
which are suitable for expressing the relative
motion of the two mesons.
It is therefore expected that this set of basis
functions is good only for the two-meson bound state.

The squared wave functions, $|\Phi(x,1-x)|^2$, for various
values of the fermion mass are shown in
Fig.\ \ref{relativeWF}. From this,
we have an intuitive picture that the mesons are
loosely bounded for small fermion
masses, while they are close to each other for
large fermion masses. This behavior has a
simple physical interpretation: As is seen
from Fig.\ \ref{mesonWF}, the
meson is tightly bounded for small fermion masses.
Therefore the
very weak Van der Waals force between the two mesons causes a
loosely bounded two-meson
state. For large fermion masses, on the other hand,
the meson has a broad shape.
Therefore the charge distribution over a wide
region keeps the two
mesons close to each other by the Coulomb force.

It is possible to check quantitatively how good the assumption
(\ref{twomeson}) is. By inspection, we find that the
wave function $\varphi$ may be written as follows,
\begin{equation}
	\varphi(x_1,x_2)=
	b_0\left({x_2\over x_1}\right)^{\beta}\ ,\label{varphi}
\end{equation}
where the constant $b_0$ is very close to $1$ for small fermion
masses. This is consistent with the massless limit
(\ref{masslessmeson}),
in which $\beta=0$, $b_0=1$, and $\varphi=1$.
Given the form of $\varphi$ (\ref{varphi}) one may
express $\Psi_2$ in terms of $\Phi$, $\psi$, and $\varphi$ as
\begin{eqnarray}
	\Psi_2(x_1,x_2)&=&
	{1\over\sqrt{2}}\int^{x_1}_0{dy\over\sqrt{y(1-y)}}
	\Phi(y,1-y)\varphi(x_1,x_1-y)\psi(x_1-y,x_2) \\
	&-&{1\over\sqrt{2}}\int^{x_2}_0{dy\over\sqrt{y(1-y)}}
	\Phi(y,1-y)\varphi(x_2,x_2-y)\psi(x_1,x_2-y)\ .
	\nonumber
\end{eqnarray}
By substituting $\psi$, (\ref{varphi}), and (\ref{relwf}),
we obtain the
following expression for $\Psi_2$ in terms of $B_l$,
\begin{eqnarray}
	\Psi_2(x_1,x_2)&=&
	{a_0b_0\over\sqrt{2}}(x_1x_2)^\beta(x_1-x_2)
	\sum^{N}_{l=0}B_l
	\sum^l_{m=0}(-1)^m\pmatrix{l\cr m}B(l+m+1,2\beta+1)
	\label{consistency}\\
	&\times&\sum^{\left[{l+m\over2}\right]}_{k=0}(-1)^k
	\pmatrix{l+m-k\cr k}(x_1x_2)^k\ ,\nonumber
\end{eqnarray}
where we used the formula,
\begin{equation}
	\sum^n_{l=0}x_1^lx_2^{n-l}=
	\sum^{\left[{n\over2}\right]}_{m=0}(-1)^m
	\pmatrix{n-m\cr m}(x_1x_2)^m\ ,
	\mbox{\rm for}\ x_1+x_2=1\ .
\end{equation}
This should be compared with (\ref{reltwo}).
Fig.\ \ref{comparison} shows
the wave function $\Psi_2$ calculated by the
direct diagonalization and by
eq. (\ref{consistency}). (The coefficients
$B_1,\cdots, B_{N}$ are
obtained by the diagonalization.
It is necessary to take into account the
normalization condition,
\begin{equation}
	\int_0^1dx|\Phi(x,1-x)|^2=1\ ,
\end{equation}
to obtain $B_0$. Although $b_0$ still remains undetermined,
we simply put $b_0=1$ for the comparison.)
The agreement measures the validity of the
concept of the ``relative motion'' of the two mesons
inside the two-meson
bound state.

%%%%%%%%%%%%%%%%%%%%%%%%%%%%%%%%%%%%%%%%%%%%%%%%%%%%%%%%%%%%%%%%
% section discussions
%%%%%%%%%%%%%%%%%%%%%%%%%%%%%%%%%%%%%%%%%%%%%%%%%%%%%%%%%%%%%%%%
\section{Discussions}\label{summary}
By using a simpler set of basis functions,
we have obtained the mass
spectrum of the massive Schwinger model in the
LFTD approximation. We
have confirmed that the two-meson bound state
has a negligibly small
six-body component and found a candidate for
the three-meson bound state. We emphasize
that the information on the wave functions is
very useful and is used for
identifying the three-meson bound state. It is
also used for
investigating the two-meson bound state. We introduce an
(approximate) meson creation operator and the concept of
the relative
motion of the two mesons. This description gives an
intuitive,
qualitative picture of the two-meson bound state
and motivates a very
simple set of basis functions.

The candidate for the three-meson bound state has
a small six-body
component compared to the two-body and four-body
components, even in the
strong coupling region. One might suspect that
it is a four-body state,
not a six-body state. Actually, the corresponding
state appears below the
threshold in the four-body calculations.
Nevertheless, we think that it
is the three-meson bound state for the following reasons.
(1) In the massless theory, there exists only a free
scalar particle in the physical
spectrum. But because the creation operator of the meson,
if expressed in
terms of the creation and annihilation operators
of the fermion and the
antifermion as in (\ref{masslessmeson}),
contains the annihilation operators, even a pure
(free) three-meson state
is not a pure six-body state. It is thus not so
strange even if it has a
small six-body component in the massive case.
(2) The state is charge
conjugation odd. A two-meson state should be
charge conjugation even. It
is hard to imagine that a two-meson state can
be charge conjugation odd,
if we rely on the description in terms of the meson.
The description may be
justified in the strong coupling region because the
unperturbed (massless)
theory has only the meson and the perturbation is small.
It is natural to
have a picture that the perturbation causes weak
interactions between the
mesons to form bound states.

Unfortunately, we do not know why the six-body component
of the three-meson bound state is so small. It is an outcome
of complex non-perturbative effects. An analysis similar to
that of Sec.\ \ref{WFRM} may reveal how the three-meson
bound state looks like but will not explain the smallness
of the six-body component. At this moment, we have to be
content with showing that it {\it is} the three-meson
bound state.

In the strong coupling region, the two- and three-meson
bound states
appear above the threshold. We have the prejudice that
despite our
numerical results, they are, in reality, still bound.
Probably they are
just below the threshold in this region and
approach the threshold in
the massless limit. We think that the reason
why they do not appear to be bound is
due to the limitation of our variational
calculations and numerical errors.
If one takes it for granted that
they are really bound, one may estimate
the errors in the calculations.

%It is interesting to note that the three-meson
%bound state is more
%tightly bounded by the two-meson bound state, i.e.,
%the binding energy of
%the three-meson bound state is greater than that of
%the two-meson bound
%state. Note that there exist such examples in Nature. Two
%Helium atoms form a very widely spreaded
%bound state just below the
%threshold (only its existence is
%experimentally confirmed %$-0.000830K$
%), while three Helium atoms form a compact,
%tightly bounded
%bound state. (The ground state has the
%binding energy $-0.108K$.)
%A similar phenomenon occurs for alpha particles.
%Two alpha particles do
%not form a bound state but a resonance just above
%the threshold
%($+92keV$), while three alpha particles form
%a bound state with the
%binding energy $-7.265MeV$.
\acknowledgments

The authors are grateful to colleagues for many discussions.
This work is supported by
a Grant-in-Aid for Scientific Research from the Ministry
of Education,
Science and Culture of Japan (No. 06640404).

\begin{figure}
\caption{
Two-body Tamm-Dancoff approximation.
The lightest states are shown with
the total number of basis functions.
The fermion mass is $m=0.1$. The
lowest state is the meson.
All the other states are ``spurious.''
}
\label{twoTD}
\end{figure}

\begin{figure}
\caption{
Four-body Tamm-Dancoff approximation.
The lightest states are shown with
the total number of basis functions.
The fermion mass is $m=0.1$. The
lowest state does not change at all.
The second lowest state goes down below
the two-meson threshold (the dashed line).
It is the two-meson bound state.
}
\label{fourTD}
\end{figure}

\begin{figure}
\caption{
Six-body Tamm-Dancoff approximation.
The lightest states are shown with
the total number of basis functions.
The fermion mass is $m=0.1$. The
lowest two states do not change at all.
The state shown by the line with triangle points
is the candidate for the three-meson bound state.
The three-meson
threshold is indicated by the dotted line.
}
\label{sixTD}
\end{figure}

\begin{figure}
\caption{
Fermion mass dependence of the mass eigenvalues.
The dashed and dotted
lines stand for the two-meson and
three-meson thresholds respectively.
}
\label{massTD}
\end{figure}

\begin{figure}
\caption{
Squared wave functions for the meson,
$|\psi(x,1-x)|^2$, are shown for
various values of the fermion mass.
For small masses the wave function
(in the momentum space) has little $x$ dependence,
implying that the meson
is a compact object, while for large masses
it is localized around $x=1/2$,
implying that the meson has a broad shape,
and that the fermion and the
antifermion are bound loosely.
}
\label{mesonWF}
\end{figure}

\begin{figure}
\caption{
Squared wave functions for the relative motion of
the two mesons in the
two-meson bound state, $|\Phi(x,1-x)|^2$,
are shown for various values
of the fermion mass.
For small masses the wave function (in the
momentum space) has a sharp peak at $x=1/2$,
implying that the mesons are
bound loosely, while for large masses it has a round shape,
implying that the mesons are very close to each other.
}
\label{relativeWF}
\end{figure}

\begin{figure}
\caption{
Consistency check for the ansatz.
The dashed line is the wave function
$\Psi_2(x,1-x)$ obtained by the direct diagonalization.
The solid line
is the wave function $\Psi_2(x,1-x)$ constructed
from $\Phi$ and $\Psi_4$
by using the ansatz (\protect\ref{twomeson}).
The fermion mass is set
$m=0.01$. The agreement is quite good. It
shows that our assumption for the two-meson bound state is
quantitatively justified {\it a posteriori}.
}
\label{comparison}
\end{figure}

\end{document}